%% file: main_reputation.tex
\documentclass[journal]{IEEEtran}
\usepackage{cite}
\usepackage{graphicx}
\usepackage[cmex10]{amsmath}
\usepackage{algorithmicx}
\usepackage{url}
\usepackage{array} 
\usepackage[caption=false,font=footnotesize]{subfig}

\newtheorem{proposition}{Proposition}

\usepackage[usenames]{color} 

\ifodd 0
\newcommand{\com}[1]{\textbf{\color{red}(COMMENT: #1)}} 
\newcommand{\comm}[1]{\textbf{\color{green}(#1)}} 
\newcommand{\clar}[1]{\textbf{\color{green}(NEED CLARIFICATION: #1)}}
\else

\newcommand{\com}[1]{}
\newcommand{\comm}[1]{}
\newcommand{\clar}[1]{}
\fi

\begin{document}

\title{Perceptions and Truth: A Mechanism Design Approach to Crowd-Sourcing Reputation}
\author{Parinaz~Naghizadeh,
        Mingyan~Liu\\
        Department of Electrical Engineering and Computer Science\\
				University of Michigan, Ann Arbor, Michigan, 48109-2122\\
				Email: \{naghizad, mingyan\}@umich.edu%
\thanks{A preliminary version of this work appeared in GameNets 2012.}}


\maketitle

\begin{abstract}
We consider a distributed multi-user system where individual entities possess observations or perceptions of one another, while the truth is only known to themselves, and they might have an interest in withholding or distorting the truth.  We ask the question whether it is possible for the system as a whole to arrive at the correct perceptions or assessment of all users, referred to as their {\em reputation}, by encouraging or {\em incentivizing} the users to participate in a collective effort without violating private information and self-interest. 
Two specific applications, {\em online shopping} and {\em network reputation}, are provided to motivate our study and interpret the results.  In this paper we investigate this problem using a mechanism design theoretic approach.  
We introduce a number of utility models representing users' strategic behavior, each consisting of one or both of a {\em truth} element and an {\em image} element, reflecting the user's desire to obtain an accurate view of the other and an inflated image of itself.  For each model, we either design a mechanism that achieves the optimal performance (solution to the corresponding centralized problem), or present individually rational sub-optimal solutions. In the latter case, we demonstrate that even when the centralized solution is not achievable, by using a simple  {\em punish-reward} mechanism, not only a user has the incentive to participate and provide information, but also that this information can improve the system performance. 
\end{abstract}


\IEEEpeerreviewmaketitle

\input{Intro} 
\input{Model}

\input{Model_I}

\input{Model_II}

\input{Model_IV}
\input{Model_V}

\input{Discussion} 
\input{Related}

\section{Conclusion}\label{sec:conclusion}
In this paper we studied the problem of designing reputation mechanisms that can incentivize users to participate in a collective effort of determining their quality assessment. We introduced a number of utility models representing users' strategic behavior, each consisting of one or both of a truth element and an image  element, reflecting the user's desire to obtain an accurate view of the other and an inflated image of itself. We demonstrated the feasibility of achieving socially optimal solutions under various combinations of user utility types by incentivizing accurate or useful input in a direct mechanism. We also presented suboptimal mechanisms when the environment is such that it is infeasible to achieve the globally optimal solution.

%


\bibliographystyle{abbrv}
\bibliography{col_rev}
%
%

%
%

\end{document}

%% file: Intro.tex
%
\section{Introduction} \label{sec:intro}

We consider a distributed multi-user system where individual entities possess observations or perceptions of one another, while the truth is only known to themselves, and they might have an interest in withholding or distorting the truth.  We ask the question whether it is possible for the system as a whole to arrive at the correct perceptions or assessment of all users, referred to as their {\em reputation}, by encouraging or {\em incentivizing} the users to participate in a collective effort without violating private information and self-interest. 
In this paper we investigate this problem using a mechanism design theoretic approach \cite{micro, game}.  We will construct a sequence of mechanisms and examine whether under each a user has incentive to participate, and if they do what they would provide as input, and whether ultimately their participation benefits the system's (global) assessment of all individuals. 

While there are various possible applications of the above problem, for the sake of concreteness we will focus on two specific application instances to motivate our study as well as to provide context within which our results can be interpreted. 



The first application is a class of online trading or shopping communities, where rating and reputation systems are routinely used, e.g., eBay, Amazon, etc \cite{josurvey}.  In this case, a buyer forms an opinion about a seller through its interactions with the latter, based on information such as 
the quality and authenticity of the received product, timeliness of the shipping process,  quality of the packaging, post-sales customer support, and so on.  The buyer has the option to provide feedback/recommendation to the site about the seller (a seller may also be able to rate a buyer on promptness of payment); two buyers' view of the same seller can differ depending on their respective experiences.  A final reputation (e.g., in the form of a score between 0 and 5) is calculated centrally by the site, e.g., taking the sum of positive feedback minus the sum of negative feedbacks, or through other similar methods \cite{josurvey, ebay02}. It is also common for the site to display the entire distribution/histograms of user input. 

Despite their simplicity and relative effectiveness, such methods have drawbacks, including the fact that not all buyers provide feedback, and those who do may have a bias, e.g., toward positive ratings \cite{josurvey, ebay02}. On the other hand, the sellers are aware of the specifics of all of their transactions (true quality).  This is an aspect that we seek to exploit in this paper in contrast to prior work. Specifically, the crowd-sourcing mechanisms developed in this paper are designed on the principle of {\em incentivizing} and collecting inputs from {\em both} buyers and sellers, on both themselves and others, with the end goal of improving the accuracy of such reputation systems.   As we shall see there exist mechanisms whereby it is in the interest of the sellers to provide {\em useful}, if not entirely {\em truthful} input. 

Our second application scenario comes from the use of Internet {\em host reputation block lists} (RBLs).  These lists are constructed by a variety of systems developed to determine the trustworthiness of a host by monitoring different types of data for suspicious behavior.  Examples of such systems include unsolicited bulk email (SPAM) lists \cite{spamhaus,spamcop}, darknet monitors \cite{darknet}, DNS sensors \cite{dnssensor}, scanning detection, firewall logs \cite{firewalllogs}, web access logs, and ssh brute force attack reports.  These lists are commonly used by network administrators to configure filters or access control lists to control incoming and outgoing traffic.
At present such data is collected and the resulting reputation lists are constructed by a handful of organizations in a way that often lacks transparency. 
It is however not hard to envision the establishment of a central system where such reputation data can be provided by participating users much as in the online shopping setting.  Peer networks naturally possess observations of each other through monitoring incoming and outgoing traffic, and may be incentivized to provide input, so that collectively the system may reach a more accurate assessment on the ``wellness'' (a measure of trustworthiness, performance, etc.) of each participant.  In this context a participant in this system can be a host or a network; in the latter case the resulting reputation refers to the quality of a network (e.g., some form of aggregated reputation of hosts in that network).   

Both the above applications, henceforth referred to as {\em online shopping} and {\em network reputation}, respectively, share the following common features.  A user in such a system can collect statistics from its interactions with other users.  From these statistics it can form certain opinions about the quality or trustworthiness of these other users, and its subsequent actions may be taken based on such opinions.  For instance, a user may choose to limit future interactions with users who have not had a satisfactory performance in the past.  Such peer user-user observations are often incomplete -- a user does not get to see the entire action profile of another user -- and can be biased.  Thus two users' view of a common third user may or may not be consistent.  
The true quality or nature of a user ultimately can only be known to that user itself (though it is possible that a user may not have this knowledge due to resource constraints). 
It is generally not in the user's self-interest to truthfully disclose this information: a user may wish to inflate others' perceptions about itself for obvious reasons: a perceived high quality, or a better \emph{public image} typically translates into other more tangible benefits, e.g., higher sales for a seller in the online shopping application and better visibility and reachability for a network in our second application. 
Similarly, a user may or may not wish to disclose truthfully what it observes about others for a variety of  considerations.  On the other hand, it is typically in the interest of a user to {\em acquire} the correct perceptions about other users.  This is because this correct view of others can help the user determine appropriate actions, e.g., a buyer naturally would like to conduct business with a trustworthy seller, while a network needs to have the correct assessment of other networks' quality in order to make effective filter configuration, routing and peering decisions, and so on. 


The design and analysis of a reputation system to be used in the above applications must observe two key features.  The first is that participation in such a system is completely voluntary, and therefore it is critical for the system to adopt mechanisms that can {\em incentivize} users to participate.  The second is that users may not report truthfully to the reputation agent even if they choose to participate in such a collaborative effort, and therefore it is crucial for any mechanism adopted by the system to either provide the right incentive to induce truthful revelation, or be able to function despite untruthful input.  These two features set the present study apart from existing work on reputation systems, many of which take user participation as a given (see e.g., the use of reputation in peer-to-peer (P2P) systems).  A detailed discussion on related work and its relationship to the present paper is given in Section \ref{sec:related}.

The rest of the paper is organized as follows. In Section \ref{sec:model}, we present the reputation system model, different elements of user utilities and some preliminaries. We present candidate mechanisms for several environments of different user types in Sections \ref{ModelI}-\ref{ModelV}. 
In Section \ref{sec:disc} we discuss main insights from these mechanisms as well as a few practical implementation issues. We review the literature of mechanism design, elicitation methods, and reputation systems most relevant to the present paper in Section \ref{sec:related}, and conclude in Section \ref{sec:conclusion}. 

%

%% file: Model.tex
\section{Model and Preliminaries} \label{sec:model}

\subsection{The reputation system model}
Consider a collection of {$K\geq 2$} entities\footnote{ We will subsequently use the terms {\em users}, {\em entities} and {\em participants} interchangeably. }, denoted by $N_1, N_2, \ldots, N_K$. A user $N_i$ may refer to a buyer/seller in an online shopping website, or a network in a system of inter-connected networks. Each user $N_i$'s overall quality is described by a quantity $r_{ii}$, which we refer to as the \emph{real} or \emph{true} quality of $N_i$, or simply the \emph{truth}. We assume without loss of generality that $r_{ii}\in [0,1], \text{for all } i=1, 2, \ldots, K$. 
We assume that each user $N_i$ is aware of its own conditions and therefore knows $r_{ii}$ precisely, but this is its {\em private information}.  We do note however that while it is technically feasible for any entity to obtain $r_{ii}$ by monitoring its own actions/interactions (e.g. a seller can record the specifics of all its transactions, a network can monitor its hosts and all outgoing traffic), it is by no means always the case due to reasons such as resource constraints. 

There is a central \emph{reputation system} that is responsible for soliciting input from participants and coming up with the system estimates. 
For instance, this could be the shopping site in the online example and a certain commonly agreed authority in the network reputation example.  Specifically, the system proposes a mechanism, according to which it collects input from participants and uses it to build a global quality assessment, in the form of a \emph{reputation index}, for each of the $K$ users in the system. Its goal is to have the  reputation index reflect the true quality  $r_{ii}$ as accurately as possible. 

In general, each user $N_j$ independently monitors its interactions with another user $N_i$ to form an estimate $R_{ji}$ based on its observations. For example, a buyer may keep track of their shopping experience with seller $N_i$. Similarly, a network $N_j$ can monitor the inbound traffic from network $N_i$ to form an opinion. However, $N_j$'s observation is in general an {\em incomplete} view of $N_i$, and may contain error depending on the monitoring and estimation technique used.  We will thus assume that $R_{ji}$ is described by a Normal distribution ${\cal N}(\mu_{ji}, \sigma^2_{ji})$, which itself may be unbiased ($\mu_{ji} = r_{ii}$) or biased ($\mu_{ji}\neq r_{ii}$) (The assumption of a Normal distribution is made for simplicity and concreteness, and is not necessary for all of our results.). 
We will further assume that this distribution is known to user $N_i$ but not necessarily to $N_j$ (i.e., it is known to the observed but not the observer),  
the reason being that $N_i$ can in principle closely monitor its own actions/interactions with $N_j$ 
and therefore may sufficiently infer how it is perceived by others. 

The reputation system itself may also be able to monitor the actions of each user $N_i$ so as to form its own estimate of $N_i$'s condition.  This will be denoted by $R_{0i}$, again a random variable for the same reason given above.  The system's observations can be gathered by a random auditing of transactions in an online shopping community, or by monitoring the outgoing traffic of a network in the network reputation example.  
As before we will assume that $R_{0i}$ is Normally distributed with ${\cal N}(\mu_{0i}, \sigma^2_{0i})$,  
%
and that this distribution is known to user $N_i$.   

The reputation system operates as follows.  It may collect a vector $(x_{ij})_{j\in K}$ of {\em reports} from each user $N_i$. It consists of \emph{cross-reports} $x_{ij}$, $i, j = 1, \cdots, K$, $j\neq i$, which represent $N_i$'s assessment of $N_j$'s quality, and \emph{self-reports} $x_{ii}, i = 1, 2, \cdots, K$, which are the users' {\em self-advertised} quality measure. As we will see in subsequent sections the mechanism may be such that only a subset of these reports are collected.  Furthermore, there is no a priori guarantee that the participants will report truthfully any of these quantities. 


The reputation system's goal 
is to derive the reputation index for each user $N_i$ so as to accurately reflect the true quality  $r_{ii}$.  This objective is quite different from what's commonly studied, e.g., revenue maximization.  Toward this end, we consider two possible ways of defining a reputation index: (1) an \emph{absolute} index $\hat{r}^A_i$ as an estimate of $r_{ii}$, and (2) a \emph{proportional} or {\em relative} index $\hat{r}^R_i$.  For instance, given true quantities $r_{ii}, i=1, 2, \cdots, K$, ideally $N_i$'s proportional reputation index is given by $\frac{r_{ii}}{\sum_{k} {r_{kk}}}$.  

Mathematically, the reputation mechanism is designed to solve the following problem: 
\begin{eqnarray}
\min \sum_{i} |\hat{r}^A_i-r_{ii}| ~\mbox{or}~ \min \sum_{i} |\hat{r}^R_i-\frac{r_{ii}}{\sum_k r_{kk}}| ~. 
\label{eq:sys_goal}
\end{eqnarray}  
Here we have used the absolute error as a performance measure; other error functions may be adopted as well. This will not change most of our subsequent analysis. 

To highlight the difference between these two types of indices, note that when using absolute reputation indices, each user's final reputation $\hat{r}^A_i$ is independent of other users' quality assessment, while proportional indices create competition among participants.  Indeed under proportional indexing users may be viewed as competing for a common pool of resources (the sum total of all index values). 
Proportional indexing in effect leads to a {\em ranking} system which may be useful in some cases.  On the other hand absolute indexing may be more relevant when used by a user to regulate its interactions with another; e.g., a buyer may refrain from shopping from poorly reputed sellers on an absolute scale even though some of them may look good by comparison.  
Similarly, a network may wish to tighten its security measure against all those with indices below a threshold, which could be the whole set or an empty set, rather than those with the poorest reputation  indices by comparison. 


A {\em reputation mechanism} specifies a method used by the reputation system to compute the reputation indices, i.e., what input to solicit and how the input are used to generate output estimates. As users are entities acting in self-interest and the truth is their own private information, the key to a successful mechanism (one that attains the solution to (\ref{eq:sys_goal})) is to induce the users to provide useful, if not entirely truthful, input.  Such a reputation mechanism will also be referred to as a {\em collective revelation} mechanism, a term borrowed from \cite{col_rev}.  It is assumed that the mechanism is common knowledge among all $K$ participating users. 

In what follows we give a brief overview of the mechanism design formalism, followed by the types of utility functions representing individual users. 


\subsection{The mechanism design framework: an overview}
\label{sec:MD} 

The theory of mechanism design \cite[Ch.~23]{micro},\cite[Ch.~7]{game}, addresses the problem of choosing the rules of a game according to the preferences of a set of agents/users, so that a desirable outcome is achieved at the equilibrium points of the resulting game.  {It is typically used to solve a decentralized resource allocation problem.}  Formally, the goal of a mechanism is to achieve the solution to a \emph{centralized problem} in an informationally decentralized system. The centralized problem is described by a triple $({\cal E}, {\cal A}, \gamma)$, where:
\begin{itemize}
	\item ${\cal E}$ is the set of all possible \emph{environments} for the problem, consisting of all the information or circumstances in the model that {are uncontrolled}. In our model, an environment $e$ consists of the utility functions, the {real quality of the participants}, the number of participating networks, etc.
	\item ${\cal A}$, the \emph{allocation space}, is the space of all feasible outcomes of the game. {In our model, an allocation space may be the set of all feasible reputation index profiles of the form $\{\hat{r}^A_i \in [0, 1]\}_{i=1}^K$ in the case of absolute indexing, and possibly tax profiles of the form $\{t_i\}_{i=1}^K$ (to be detailed shortly).}
	\item $\gamma$, the \emph{goal correspondence}, is a mapping $\gamma:{\cal E}\rightarrow {\cal A}$ that achieves some desired performance goal, {e.g., the maximization of a social choice rule.}  
\end{itemize} 

In an informationally decentralized system, the mechanism designer chooses the game form $({\cal M}, h)$, where:
\begin{itemize}
	\item ${\cal M}=\Pi_{i=1}^K {\cal M}_i$, with ${\cal M}_i$ denoting the \emph{message space} of user $i$.
	\item $h:{\cal M}\rightarrow {\cal A}$ is the \emph{outcome function}. This function is the rule according to which the mechanism uses the collected input messages to compute the final allocation.
\end{itemize} 
Define $\xi({\cal M}, h, e)$ as the outcome at the equilibrium point\footnote{Note that the appropriate equilibrium concept depends on the model; we will highlight this within each model we study.} 
  of the game induced by $({\cal M}, h)$ when the realization of the environment is $e\in {\cal E}$. The game form is chosen such that $\xi({\cal M}, h, e) \subseteq \gamma(e)$. {In other words, the induced game implements in its equilibrium the solution to the centralized allocation problem.} 

In addition to implementing the desired outcome, a game form is often required to satisfy other properties. One such desirable property is \emph{budget balance}. 
{Note that in order to induce individuals to behave in such a way that the solution to the centralized problem is obtained, the mechanism typically needs some type of {\em leverage}.  The precise form of this leverage varies from problem to problem (see more discussion in Section \ref{sec:related}), but the most commonly used leverage is {\em taxation}, an amount $t_i$ imposed on user $i$: a user is taxed/punished ($t_i>0$) for bad behavior and credited/rewarded ($t_i<0$) for good behavior; a user's valuation of tax (monetary payout) is assumed public knowledge.} 
{How taxation if invoked may be implemented in our problem context is discussed in Section \ref{sec:disc}.} 
%

{A balanced budget refers to the fact that at equilibrium all money collected (tax) equals all money paid out (credit); i.e., the system running the mechanism neither profits nor subsidizes but merely uses taxation as a regulatory tool. } 
By contrast, a budget deficit implies that the system needs to inject money into the system, while a budget surplus means that some amount of money will be left unclaimed. 
For this reason, it is desirable to reallocate the paid taxes in the form of subsidies and ensure a balanced budget, i.e. to have $\sum_i t_i=0$. 

More importantly, it is desirable to design a mechanism that is \emph{individually rational}: a user benefits from participation. In other words, the expected utility from playing the game induced by the mechanism should exceed the reserved utility a user gets when staying out.  {We next introduce the types of utility functions considered in this study.} 


\subsection{Individual users' objectives}{\label {utility}} 
%
In modeling the users' objectives, we identify two elements of a user's utility or preference model. 
\begin{itemize}
	\item \emph{Truth}: Each user $N_i$ may wish to obtain from the system as accurate as possible an estimate on users $N_j$ {\em other than itself}. Formally,
\begin{eqnarray*}
I_i &=& - \sum_{j\neq i} f_i(| \hat{r}^A_j - r_{jj} |) ~, 
\\
I_i &=& - \sum_{j\neq i} f_i(| \hat{r}^R_j - \frac{r_{jj}}{\sum_{k} r_{kk}} |) ~, 
\end{eqnarray*}
for absolute and proportional/relative reputation indices, respectively. Here, $f_i(\cdot)\geq 0$ are increasing and convex functions.  

This element captures a user's interest in having accurate assessment of other users' quality so that it can properly regulate its actions.  For instance, it is important for a buyer (or trader) to know the history of a seller's (or potential partner's) transactions. Similarly, it would be important for a consumer-based network like Comcast whose customers connect to various networks/sites to have an accurate view of these other networks.  

\item \emph{Image}: Each user $N_i$ may further wish to obtain as high as possible an estimate on {\em itself}. Formally,
	\begin{eqnarray*}
II_i = g_i(\hat{r}^A_i) ~,~~~~
II_i = g_i(\hat{r}^R_i),
\end{eqnarray*}
for absolute and proportional reputation indices respectively, where $g_i(\cdot)\geq 0$ are concave and increasing. 

This element reflects a user's interest in having a high reputation itself as it translates into other tangible benefits as mentioned earlier. 
For instance, a seller is interested in attracting costumers by building a good reputation and increasing its visibility. Similarly, this objective can capture a content-centric network like Craigslist or a blog hosting site, for which staying visible to the outside world is critical and who may reserve the right to block users from certain networks.   
\end{itemize} 

A general preference model of a legitimate, non-malicious user may consist of both elements, possibly weighted;
that is, user $N_i$ may be captured by 
\begin{eqnarray}
u_i = - \lambda \sum_{j\neq i} f_i(| \hat{r}^A_j - r_{jj} |) + (1-\lambda)g_i(\hat{r}^A_i)
\end{eqnarray} 
for some constant $0\leq\lambda\leq 1$ (and similarly for relative reputation indices). 

It is interesting to note the two extreme cases: (1) There are users that are only concerned with truth ($\lambda= 1$), e.g. a buyer on Amazon, or a closed network such as a DoD network that has strict requirement on which external sites it is allowed to connect to but does not care about its own image to the outside world. These users will be referred to as the {\em truth type}. (2) There are also users that are only concerned with their images ($\lambda=0$), e.g. a seller on Amazon, or a phishing site that tries to maximize its reputation in order to attract more traffic. These users will be referred to as the {\em image type}. The more general model that consists of both will be referred to as a {\em mixed type}. Examples of mixed type users include a user in an online trading community; or a seller on eBay, who in addition to gaining popularity by a high reputation, benefits from using the knowledge of the reputation of buyers to ensure receiving the promised payments. 
With the above classifications, depending on the makeup of the system, we may have a \emph{homogeneous} environment where all participants are of the same type, or a \emph{heterogeneous} environment with a mixture of different types. 

By defining these two utility elements, we assume a user's preference is {in general} increasing in the accuracy of others' quality estimate, and increasing in its own quality estimate. We assume these two characteristics to be public knowledge. How the preference increases with these estimates and how these two elements are weighed, i.e. the functional forms of $f_i(\cdot)$ and $g_i(\cdot)$, remain user $N_i$'s private information in general. 

It should be noted that $\hat{r}^A_j$ (and similarly $\hat{r}^R_j$) is a function of the proposed game form $({\cal M}, h^A)$, such that, $\hat{r}^A_j = h^A({m}_1, {m}_2, \ldots, {m}_K)$, with $m_j$ denoting $N_j$'s message. Since the proposed model is one of incomplete information, from $N_i$'s viewpoint, the message profile ${m}\in {\cal M}$, and consequently $u_i$, is in general a random variable. Therefore, it is understood that $N_i$ is an \emph{expected}-utility maximizer. 
Also, if a user $N_i$ is charged a tax in the amount $t_i$ according to the specific mechanism, then  $N_i$'s aggregate utility is given by $v_i := u_i - t_i$. 

{Note that the utility model assumed above may not capture the nature of a {\em malicious} user, who may or may not care about the estimated perceptions about itself or others.  This is discussed further in Section \ref{sec:disc}.} 
\subsection{Solution to the centralized problem} 

For the centralized problem given in (\ref{eq:sys_goal}), if 
the reputation system has full information about all the parameters in the system, then the optimal choice of the absolute and proportional reputation indices would simply be $\hat{r}^A_i=r_{ii}$  and $\hat{r}^R_i=\frac{r_{ii}}{\sum_{k} {r_{kk}}}$, for $i=1,2,\ldots, K$, respectively. 

In subsequent sections we show how to design a reputation mechanism {in a decentralized scenario}, {for various combinations of the utility elements} described in Section \ref{utility}, such that the centralized solution is an equilibrium of the resulting game when played by the users, or to find a suboptimal mechanism when the centralized solution cannot be implemented. 

%% file: Model_I.tex
\section{Truth type, absolute reputation} \label{ModelI}

{Our first case deals with a homogeneous environment in which all users are of the truth type, with the following utility function: 
\begin{eqnarray}
\mbox{(Model I)} & u_i =& - \sum_{j\neq i} f_i(| \hat{r}^A_j - r_{jj} |) ~. 
\label{eq:ind_goal_Ia}
\end{eqnarray}  }
{Below we first present a mechanism that can achieve the centralized solution $\hat{r}^A_i=r_{ii}$, and then discuss the properties of the game it induces.} 

\subsection{The Absolute Scoring Mechanism}
The Absolute Scoring (AS) mechanism consists of the following components: 
\begin{itemize}
	\item Message space ${\cal M}$: each user reports a single value $x_{ii}\in [0,1]$ as its message.
	\item Outcome function $h(\cdot)$: 
		The reputation system sets the reputation indices $\hat{r}^A_i=x_{ii}, \forall i$.
		
	\item In addition, user $N_i$ is levied a tax in the amount $t_i$ based on its own report $x_{ii}$, other reports $x_{jj}$, $j\neq i$, and the system's own observation $R_{0i}$:
		\begin{eqnarray}
		t_i = | x_{ii}-R_{0i}|^2 -\frac{1}{K-1} \sum_{j \neq i} | x_{jj}-R_{0j}|^2.
		\label{eq:taxI}
		\end{eqnarray}
\end{itemize}
The rationale behind this mechanism is as follows: the system assigns the reputation assuming truthful reports; at the same time it ensures that the reports are indeed truthful by choosing the appropriate format for the tax transfer, partly utilizing its own knowledge.

\subsection{Properties of the AS mechanism} 
As discussed earlier, our objective in designing a mechanism is to design a game form that  implements the centralized solution. Therefore, we first verify that truth-telling is an equilibrium of the induced game.  Under our model the resulting game is a game of incomplete information: each user only knows its own environment and has a belief (i.e. a probability distribution) on the set of environments of all participating agents (i.e. the other users as well as the reputation system). Therefore, truth-telling should be implemented in either Bayesian Nash equilibrium or dominant strategies. 

\begin{proposition}
Truth-telling is a dominant strategy in the game induced by the Absolute Scoring mechanism.
\end{proposition}
\begin{IEEEproof}
We need to show that $N_i$'s expected utility when choosing the message $x_{ii}$ 
under any strategy profile $\{x_{jj}\}_ {j\neq i}$ for all other users, is maximized at $x_{ii}=r_{ii}$. $N_i$'s expected utility is given by: 
\begin{eqnarray}
E\left[ v_i (x_{ii}, \{x_{jj}\}_{j\neq i})\right] =
&-& \sum_{j\neq i} E[f_i(| \hat{r}^A_j - r_{jj} |)]\notag\\
- E[\ | x_{ii}-R_{0i}|^2] &+& \tfrac{1}{K-1} \sum_{j \neq i} E[| x_{jj}-R_{0j}|^2]~~~~
\label{eq:u_Ia}
\end{eqnarray}It can be easily seen that $N_i$'s report $x_{ii}$ can only adjust the second term in \eqref{eq:u_Ia} regardless of other participants' strategies. $N_i$ is a self-utility maximizer, therefore $x_{ii}$ is chosen so as to minimize the second term, i.e., minimize the punishment due to discrepancy w.r.t to the system's observation. By assumption, $N_i$ knows that $R_{0i}\sim {\cal N}(r_{ii}, \sigma_{0i}^2)$\footnote{Assuming unbiased observations. If cross-observations are biased, the analysis will depend on which entities have a knowledge about the existence and/or distribution of such bias. }, and it is easy to see that the optimal choice is indeed $x_{ii}=r_{ii}$.  
\end{IEEEproof}

In addition to implementing the centralized solution in dominant strategies, the AS mechanism is also budget balanced and individually rational. 
To see the first property, note that by \eqref{eq:taxI} 
the system is simply charging each user by their inaccuracy as compared to its own observation, and then redistributing the gathered fees among the other participants. 

Before addressing the individual rationality requirement, we note that as described in Model I, $N_i$ is interested in a vector of reputation indices that will help it regulate its interactions with other users $N_j,\ j\neq i$. Therefore, our argument depends on the environment ${\cal E}_i$, i.e., the information that $N_i$ initially holds about the system's parameters.
\begin{proposition}
The AS mechanism is individually rational for all possible environments $e\in {\cal E}$.
\end{proposition}
\begin{IEEEproof}
	As discussed in Section \ref{sec:model}, a user $N_i$ is able to obtain its own observation $R_{ij}$ on $N_j$'s true quality, 
	such that the best estimate $N_i$ can form on $r_{jj}$ is an unbiased observation $R_{ij}\sim {\cal N}(r_{jj}, \sigma_{ij}^2)$. 
	%
	Consequently, its expected reserved utility is given by $- \sum_{j\neq i} E(f_i(| R_{ij} - r_{jj} |))$ if it chooses to stay out of the system. On the other hand, participation will result in an expected utility of $- \sum_{j\neq i} f_i(0)$ at equilibrium. Since by assumption $f_i(\cdot)$ is convex, we have, $\forall j\neq i$,
	\begin{eqnarray}
	E[f_i(| R_{ij} - r_{jj} |)] &\geq& f_i(E(| R_{ij} - r_{jj} |)) \nonumber \\
	&=& f_i(\sqrt{\frac{2}{\pi}}\sigma_{ij}) \nonumber \\
	&>& f_i(0) \ \forall j\neq i ~,
	\end{eqnarray} 
where the last inequality is due to the fact $f_i(\cdot)$ is an increasing function. 
Thus 
	the proposed mechanism is individually rational.
\end{IEEEproof}


It is interesting to note that in this scenario, the solution $\hat{r}_i^A = r_{ii}$ is both socially and individually optimal. Therefore, it should come as no surprise that the AS mechanism manages to implement the socially optimal solution while being incentive compatible, individually rational, and budget balanced.  

\subsection{The Extended-AS Mechanism}
We end this section by presenting an extension to the AS mechanism when the reputation system does not possess direct observations. Specifically, the reputation system adopts a random \emph{ring}, which may be made explicit to the users or kept secret from them (some implications of this choice are discussed in section \ref{collusion}). 
 
Assume users are re-labeled according to their positions on this ring, such that $N_{i+1}$ follows $N_i$ and so on. The self-report of a user is then validated using the cross-report from its predecessor on the ring. More specifically, each user $N_i$ is asked to provide a self-report $x_{ii}$ (which will be assigned as its reputation index $\hat{r}_i^A$), and cross-reports $x_{ij}$, only one of which -- $x_{i(i+1)}$ for its successor $N_{i+1}$ -- is used in the mechanism. $N_i$ is then levied a tax based on the discrepancy between its self-report and the cross-report $x_{(i-1)i}$, given by: 
\begin{eqnarray*}
t_i &=& | x_{ii}-x_{(i-1)i}| - \frac{1}{K-2} \sum_{j \neq i, i+1} | x_{jj}-x_{(j-1)j}|~.
\end{eqnarray*} 
The summation term in $t_i$ is a share of taxes collected from users other than $N_i$ and its immediate neighbor, ensuring budget balance. 

We now verify that these tax terms lead to truthful self-reports and cross-reports. Furthermore, the mechanism is individually rational, and implements the centralized solution in Bayesian Nash equilibrium. 
Note that the aggregate utility of $N_i$ in the extended-AS mechanism is given by: 
\begin{eqnarray*}
v_i  = - \sum_{j \neq i, i+1} f_i(|x_{jj} - r_{jj} |) - f_{i}(\left|x_{(i+1)(i+1)} - r_{(i+1)(i+1)}\right|)\\
     - | x_{ii}-x_{(i-1)i}| +\ \frac{1}{K-2} \sum_{j \neq i, i+1} | x_{jj}-x_{(j-1)j}|~.
\end{eqnarray*} 

\begin{proposition} The Extended-AS mechanism results in truthful self-reports and cross-reports. 
\end{proposition}
\begin{IEEEproof}
	First, let's consider the self-report $x_{ii}$: this report appears only in the term $| x_{ii}-x_{(i-1)i}|$. To minimize the expected value of this term (maximize $E[v_i]$), $N_i$ chooses $x_{ii}= E[x_{(i-1)i}]$. Therefore if the cross-report $x_{(i-1)i}$ is truthful, $N_i$ will provide the truthful self-report $x_{ii}=r_{ii}$. 
	
	Following the previous argument, $N_{i+1}$ will also choose $x_{(i+1)(i+1)} = \arg\min_{x} E(|x-x_{i(i+1)}|)\ $. $N_{i}$ submits the unbiased $R_{i(i+1)}$, by predicting that a truthful self-report is a best-response to \emph{unbiased} cross-observations. This incentivizes $N_{i+1}$ to reveal $r_{(i+1)(i+1)}$. 
\end{IEEEproof}

We conclude that the centralized solution $\hat{r}_i^A=r_{ii}$ is implemented in a Bayesian Nash equilibrium. Individual rationality follows from Proposition 2, since the expected value of the tax term is zero.

%% file: Model_II.tex
\section{Truth type, relative reputation} \label{ModelII}

We now turn to the case where the reputation system seeks to calculate relative/proportional indices for a homogeneous environment consisting of $K$ truth type participants, with utility functions given by: 
\begin{eqnarray}
\mbox{(Model II)} & u_i =& - \sum_{j\neq i} f_i(| \hat{r}^R_j - \frac{r_{jj}}{\sum_{k}{r_{kk}}} |)~. 
\label{eq:ind_goal_Ib}
\end{eqnarray} 
Consider the following \emph{Fair Ranking} (FR) mechanism: 
\begin{itemize}
	\item Message space ${\cal M}$: each user reports one value $x_{ii}\in [0,1]$ as its self-advertised reputation.
	\item The outcome function $h(\cdot)$: the system assigns the proportional reputations $\hat{r}^R_i = \frac{x_{ii}}{\sum_{k}{x_{kk}}}$. No taxes are assessed. 
\end{itemize} 
It turns out the above mechanism achieves the centralized solution $\frac{r_{ii}}{\sum_{k}{r_{kk}}}$ as stated formally in the next proposition. 
In essence, this is an incentive compatible {\em direct mechanism} in which truthfully reporting the real quality is a Bayesian Nash equilibrium. 
\begin{proposition}
Truthful revelation is a Bayesian Nash equilibrium of the Fair Ranking mechanism. 
\end{proposition}
\begin{IEEEproof}
Consider $N_i$'s utility from reporting $x_{ii}$ when all other users are truthfully disclosing their real quality $r_{kk},\ k\neq i$. We have:
\begin{eqnarray}
u_i(x_{ii}, \{r_{kk}\}_{k\neq i}) = - \sum_{j\neq i} f_i(|\tfrac{r_{jj}(x_{ii}-r_{ii})}{(x_{ii}+\sum_{k\neq i} r_{kk})(\sum_k r_{kk})}|).
\label{eq:model3}
\end{eqnarray}
By assumption, $f_i(\cdot)$ is an increasing function, and therefore $N_i$'s best response is to report $x_{ii}=r_{ii}$. Note that this result is achievable without the need for cross-observations from other users, direct observations by the system itself, or taxation. 
\end{IEEEproof} 

It is interesting to highlight the difference between Models I and II. While both models are based on the same utility type (truth) and the centralized, full information solution is implementable in both cases, under Model II the mechanism induces truth-telling without the need to impose taxes. 
This is due to the fact that with proportional indices, a user's report influences other users' allocation, an effect missing in the case of absolute indices. The outcome obtained as a result of this effect is rather intuitive: when all individuals are interested in establishing a fair \emph{ranking} system (Model II), 
truthful revelation is the best strategy for all. 

%% file: Model_IV.tex
\section{Mixed type, absolute reputation}\label{ModelIV}




Our last {homogeneous} case deals with the mixed type with absolute reputation given by the following utility function: 
\begin{eqnarray}
\mbox{(Model III)} ~ u_i = - \sum_{j\neq i} f_i(| \hat{r}^A_j - r_{jj} |) + g_i(\hat{r}^A_i)~. 
\label{eq:ind_goal_IIa}
\end{eqnarray}

\subsection{An impossibility result} 

As in the case of Model I noted earlier, Model III also leads to a game of incomplete information, therefore the appropriate equilibrium concept is that of Bayesian Nash equilibrium. A necessary condition for a goal correspondence to be truthfully implementable in Bayesian Nash equilibrium in an economic environment, is Bayesian incentive compatibility of that goal correspondence \cite{jackson91}, \cite[Ch.~23]{micro}, defined as follows {(recall the goal correspondence $\gamma$ achieving the centralized solution is $\gamma(e) = \{{r_{ii}}\}_{i=1}^{K}$)}: 

A social choice correspondence $\gamma: {\cal E} \rightarrow {\cal A}$ is Bayesian incentive compatible if and only if for every $i \in K$,
\begin{eqnarray}
\int_{{\cal E}_{-i}} u_i(\gamma(e_i,e_{-i}))p(e_{-i}|e_{i})\mathrm{d}e_{-i}\notag\\
\geq \int_{{\cal E}_{-i}} u_i(\gamma(e'_i,e_{-i}))p(e_{-i}|e_{i})\mathrm{d}e_{-i} ~.
\label{eq:BIC}
\end{eqnarray} 

This means that (for our model) any social choice function that is not Bayesian incentive compatible cannot be implemented in BNE.  
{Below we show that} the desired goal correspondence for Model III does not satisfy \eqref{eq:BIC}, thus there is no game form that can achieve the allocation $\hat{r}^A_i=r_{ii}$, and that the centralized solution is not implementable under this model. 

Assume the realized environment of user $N_i$ is $e_i\in {\cal E}_i$, and according to this environment, the true quality of $N_i$ is $r_{ii}$. Therefore, the resulting optimal solution as prescribed by $\gamma(\cdot)$ is to have $\gamma(e_i,e_{-i})=(r_{ii}, \{r_{jj}\}_{j\neq i})$. If however, $N_i$ misrepresents its environment by claiming the true quality is $r'_{ii}>r_{ii}$, the allocation would be $\gamma(e_i,e_{-i})=(r'_{ii}, \{r_{jj}\}_{j\neq i})$.  
{This change does not affect  
the first term in {the utility function}, while causing  
the second term 
to increase since by assumption $g_i(\cdot)$ is an increasing function. This in turn means (\ref{eq:BIC}) does not hold. 
In other words, any mechanism under this model inevitably has some performance gap  compared to the centralized solution in terms of its mean absolute error (MAE) as given in the centralized objective function.

\subsection{The use of self-reports and cross-reports} 

In view of the impossibility result, we next set out to construct a good, suboptimal mechanism.  We will invoke the use of both self-reports and cross-reports, and will forgo the use of taxation for simplicity.  As we shall see later, even though the system's own observation $R_{0i}$ is sufficient in implementing our mechanism, more cross-reports can improve the performance of the mechanism when used properly. 
 
We first introduce a simple, benchmark mechanism, referred to as the {\em simple averaging mechanism}, where the reputation agent solicits cross-reports $x_{ji}$, and computes the estimate $\hat{r}^A_i$ as the average of $x_{ji}$ for $j\neq i$ and its own observation $R_{0i}$. %
This is the basic mechanism used in many existing online systems, e.g., Amazon and Epinions \cite{josurvey}.  
The following proposition shows that for this mechanism, $N_j$ will choose to truthfully disclose its observation $R_{ji}$. 
\begin{proposition}
Under the simple averaging mechanism, truthful revelation of the observation $R_{ji}$, by $N_j$,  $j\neq i$ is a Bayesian Nash equilibrium.
\end{proposition}
\begin{IEEEproof}
Note that a user $N_j$ has no influence on its own estimate $\hat{r}^A_j$, which is only a function of  other users' input. Thus in effect $N_j$'s objective is to minimize the error $f_j(|\hat{r}^A_i-r_{ii}|)$ for all other network $N_i$. The simple averaging mechanism will result in $\hat{r}^A_i=\frac{\sum_{k \in K\backslash i} R_{ki} + R_{0i}}{K}$, which for unbiased estimates $R_{ki}$, will also be an unbiased estimator of $r_{ii}$, i.e. $\hat{r}^A_i\sim {\cal N}(r_{ii}, \frac{\sigma^2}{K})$. Any deviation by $N_j$ will shift the mean of $\hat{r}^A_i$, thus degrading the estimate and increasing the error.
\end{IEEEproof}

If the estimates $R_{ji}$, for $j\in K\backslash i$, are unbiased, then $\hat{r}^A_i$ can be made arbitrarily close to $r_{ii}$ as the number of participants increases. 
%
%
It's not hard to see that under this mechanism, if asked, $N_i$ will always report $x_{ii}=1$, and thus the self-reports will bear no information. 
%

Alternatively, we could seek to build a mechanism that incentivizes $N_i$ to provide a {\em useful} self-report even if it is not the precise truth $r_{ii}$. 
With this in mind, a good mechanism might on one hand convince $N_i$ that it can help contribute to a desired, high estimate $\hat{r}^A_i$ by supplying input $x_{ii}$, while on the other hand try to use the cross-reports, which are estimates of the truth $r_{ii}$, to assess $N_i$'s self-report and threaten with punishment if it is judged to be overly misleading. 

Furthermore, it is desirable for the mechanism to be such that $N_i$'s cross-reports are not used in deriving its own reputation. By doing so, we ensure that the cross-reports are reported truthfully\footnote{This is conceptually similar to not using a user's own bid in calculating the price charged to him in the context of auction, a technique commonly used to induce truthful implementation.}. To see why this is the case, note that by sending its cross-report on $N_j$, $N_i$ can now only hope to increase its utility by altering the term $f_i(|\hat{r}^A_j-r_{jj}|)$. 
$N_i$'s best estimate of $r_{jj}$ is its cross-observation $R_{ij}$, which it knows will be used as a basis for the estimate $\hat{r}^A_j$. On the other hand, due to its lack of knowledge of $r_{jj}$, $N_i$ cannot determine 
how to manipulate $x_{ij}$ so as to increase its utility. By this argument, for the rest of this section we will assume that the cross-reports are reported truthfully under the Model III utility type, and that this is common knowledge. 

It is worthwhile to emphasize that the above argument is based on the \emph{direct} effect of the cross-reports on the final reputation. One might argue that $N_i$ could exploit the \emph{indirect} effect of its cross-report by badmouthing other users so as to improve its \emph{relative} position in the system, i.e., make itself look better by comparison. However, there is no clear incentive for $N_i$ to do so, since the current model is one of absolute, rather than proportional reputations. 

But more importantly and perhaps more subtly, badmouthing another user is not necessarily in the best interest of an individual. Suppose that after sending a low cross-report $x_{ij}$, $N_i$ subsequently receives a low $\hat{r}^A_j$ from the reputation system. Due to its lack of knowledge of other users' cross-reports, $N_i$ cannot reasonably tell whether this low estimate $\hat{r}^A_j$ is a consequence of its own low cross-report, or if it is because $N_j$ was observed to be poor(er) by other users and thus $\hat{r}^A_j$ is in fact reflecting $N_j$'s true quality (unless a set of users {\em collude} and jointly target a particular individual). 
This ambiguity is against $N_i$'s interest in obtaining accurate estimates of other users; therefore bashing is not a profitable deviation from truthful reporting. 
In essence, the desire for truth (accuracy) gives the system leverage in designing a mechanism even if it's only part of the user's objective.  This is discussed further in Section \ref{sec:disc}. 



\subsection{The punish-reward (PR) mechanism}

Consider the following way of computing the reputation index $\hat{r}^A_i$ for $N_i$. The system  uses its own observation $R_{0i}$, along with the received cross-reports $R_{ji}$, to judge $N_i$'s self-report. In the simplest case, the system can take the average of all these estimations to get $\bar{x}_{0i}:= \frac{\Sigma_{j\in K\backslash i} x_{ji} + R_{0i}}{K}$, and derive $\hat{r}^A_i$ using:
\begin{align}
 \hat{r}^A_i(x_{ii}, \bar{x}_{0i}) = 
 \begin{cases}
\frac{\bar{x}_{0i}+x_{ii}}{2} & \textrm{if~} x_{ii} \in [\bar{x}_{0i}-\epsilon, \bar{x}_{0i}+\epsilon], \\
\bar{x}_{0i}-|\bar{x}_{0i}-x_{ii}| & \textrm{if~} x_{ii} \notin [\bar{x}_{0i}-\epsilon, \bar{x}_{0i}+\epsilon].
\end{cases}
\label{eq:GameForm}
\end{align}
where $\epsilon$ is a fixed and known constant.  
In words, the reputation system takes the average of the self-report $x_{ii}$ and the \emph{aggregate cross-report} $\bar{x}_{0i}$ if the two are sufficiently close, or else punishes $N_i$ for reporting significantly differently.  
We refer to this mechanism as the \emph{punish-reward} mechanism. Note that this is only one of many possibilities that reflect the idea of weighing between averaging and punishing; for instance, we can also choose to punish only when the self-report is higher than the cross-report, and so on. 

Next we examine the strategic behavior of the users when playing the induced game. Throughout the analysis, we will assume that all cross-observations are unbiased and are reported truthfully as argued in the previous sub-section,   
i.e., $x_{ji}\sim {\cal N} (r_{ii}, \sigma^2), j\in K\backslash i$ and $R_{0i}\sim {\cal N} (r_{ii}, \sigma^2)$.\footnote{Note that we are assuming $\sigma$ is common, and known by the reputation system as well as the participants. $\sigma$ can be thought of as a measure of the variation of the estimates on $N_i$, which depends on the nature of the observations and the algorithm used for the estimate. While this is not an unreasonable assumption, verification using real data is desired which is currently being pursued in a parallel effort.}  

\subsection{Value of cross-report and self-report}
Since $N_i$ knows the distribution of the observations $R_{ji}$, it will assume the aggregate cross-report is a sample of a distribution ${\cal N}(\mu, \sigma'^2)$, with $\mu=r_{ii}$ and $\sigma'^2 = \frac{\sigma^2}{K}$. The choice of the self-report $x_{ii}$ is then determined by the solution to the optimization problem $\max_{x_{ii}} E[g_i(\hat{r}^A_i)]$. 

To simplify the following calculation, we will take the special case $g_i(x)=x$. The analysis can be easily extended to other functional forms of $g_i(\cdot)$. 
Using \eqref{eq:GameForm}, $E[\hat{r}^A_i]$ eventually simplifies to (with $F()$ and $f()$ denoting the cdf and pdf of $\bar{x}_{0i}$, respectively): 
\begin{eqnarray}
E[\hat{r}^A_i] &=& x_{ii} + \tfrac{\epsilon}{2}(F(x_{ii}+\epsilon)-3F(x_{ii}-\epsilon))\notag\\
&-& \tfrac{1}{2}{\int_{x_{ii}-\epsilon}^{x_{ii}+\epsilon}} {F(x)\mathrm{d}x} - 2{\int_{-\infty}^{x_{ii}-\epsilon}} {F(x)\mathrm{d}x}~.~~ 
\label{eq:FinalOpt}
\end{eqnarray}
Taking the derivative with respect to $x_{ii}$, we get:
\begin{eqnarray}
\frac{\mathrm{d}E[\hat{r}^A_i]}{\mathrm{d}x_{ii}} &=& 1 + \frac{\epsilon}{2}[f(x_{ii}+\epsilon) - 3f(x_{ii}-\epsilon)]\notag\\
&-& \frac{1}{2}[F(x_{ii}+\epsilon)+3F(x_{ii}-\epsilon)] . 
\label{eq:Derivative}
\end{eqnarray}

We next re-write $\epsilon=a\sigma'$; this expression of $\epsilon$ reflects how the reputation system can limit the variation in the self-report using its knowledge of this variation $\sigma'$. Replacing $\epsilon=a\sigma'$ and $\bar{x}_{0i} \sim {\cal N}(\mu,\sigma'^2)$ in \eqref{eq:Derivative}, and making the change of variable $y:=\frac{x_{ii}-\mu}{a\sigma'}$ results in:
\begin{eqnarray}
\frac{a}{\sqrt{2\pi}}(e^{-(\frac{a(y+1)}{\sqrt{2}}) ^2} -3 e^{-(\frac{a(y-1)}{\sqrt{2}})^2})&&\nonumber\\
- \frac{1}{2} (\text{erf}(\frac{a(y+1)}{\sqrt{2}}) + 3\text{erf}(\frac{a(y-1)}{\sqrt{2}})) &=& 0 ~. 
\label{eq:Simplified}
\end{eqnarray}
Therefore, if $y$ solves \eqref{eq:Simplified} for a given $a$, the optimal value for $x_{ii}$ would be $x_{ii}^{*} = \mu + a\sigma' y$. Equation \eqref{eq:Simplified} can be solved numerically for $a$, resulting in Fig. \ref{y_vs_a}. 

Two interesting observations can be made  from Fig. \ref{y_vs_a}: (1) $0< y < 1$, and (2) as a consequence $\mu< x_{ii}^{*} < \mu+\epsilon$. This means that $N_i$ chooses to inflate its self-report in hope of inflating $\hat{r}^A_i$, while trying to stay within its prediction of the acceptable range.

\subsection{Properties of the PR mechanism}
We first compare the performance of \eqref{eq:GameForm} to the simple averaging mechanism. Define $e_{m}:=E[|\hat{r}^A_i-r_{ii}|]$ as the MAE of the mechanism described in \eqref{eq:GameForm} with $\epsilon=a\sigma'$. Assuming the optimal self-report $x_{ii}^*$, and unbiased, truthful cross-reports, {it is possible to find the expression for $e_m$ as a function of the parameter $a$. We can thus optimize the choice of $a$ by solving the problem $\min_{a} e_{m}$. Taking the derivative of $e_m$ we get:}  
\begin{eqnarray}
\frac{\mathrm{d}e_{m}}{\mathrm{d}a} &=& \tfrac{\sigma'}{2}\Big(\tfrac{a}{\sqrt{2\pi}}(e^{-\frac{(a(y+1))^2}{2}}-3e^{-\frac{(a(y-1))^2}{2}})
+ (ay+y')\nonumber\\
&& \big( \text{erf}(\tfrac{ay}{\sqrt{2}}) - \frac{1}{2}(\text{erf}(\tfrac{a(y+1)}{\sqrt{2}})-3\text{erf}(\tfrac{a(y-1)}{\sqrt{2}}))\notag\\
&+&
\frac{a}{\sqrt{2\pi}}(e^{-\tfrac{(a(y+1))^2}{2}}+3e^{-\tfrac{(a(y-1))^2}{2}}) + 2\big)\Big) ~. 
\label{eq:ErrorDer}
\end{eqnarray}

\begin{figure*}
\begin{minipage}[b]{0.33\linewidth}
\includegraphics[width=0.8\columnwidth]{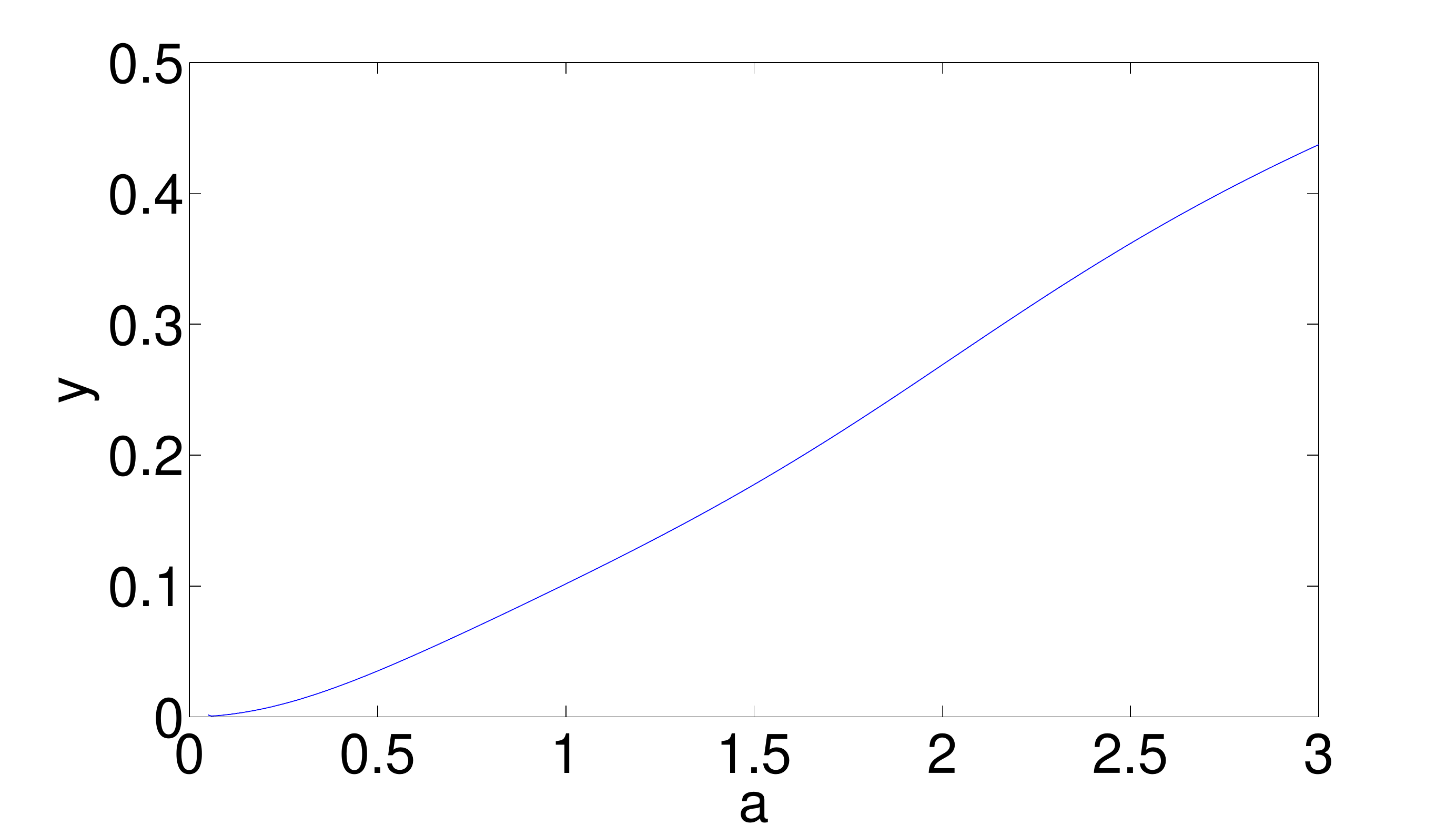}%
\caption{Solution of \eqref{eq:Simplified}: $y$ vs. $a$}%
\label{y_vs_a}%
\end{minipage}
\begin{minipage}[b]{0.33\linewidth}
\includegraphics[width=0.8\columnwidth]{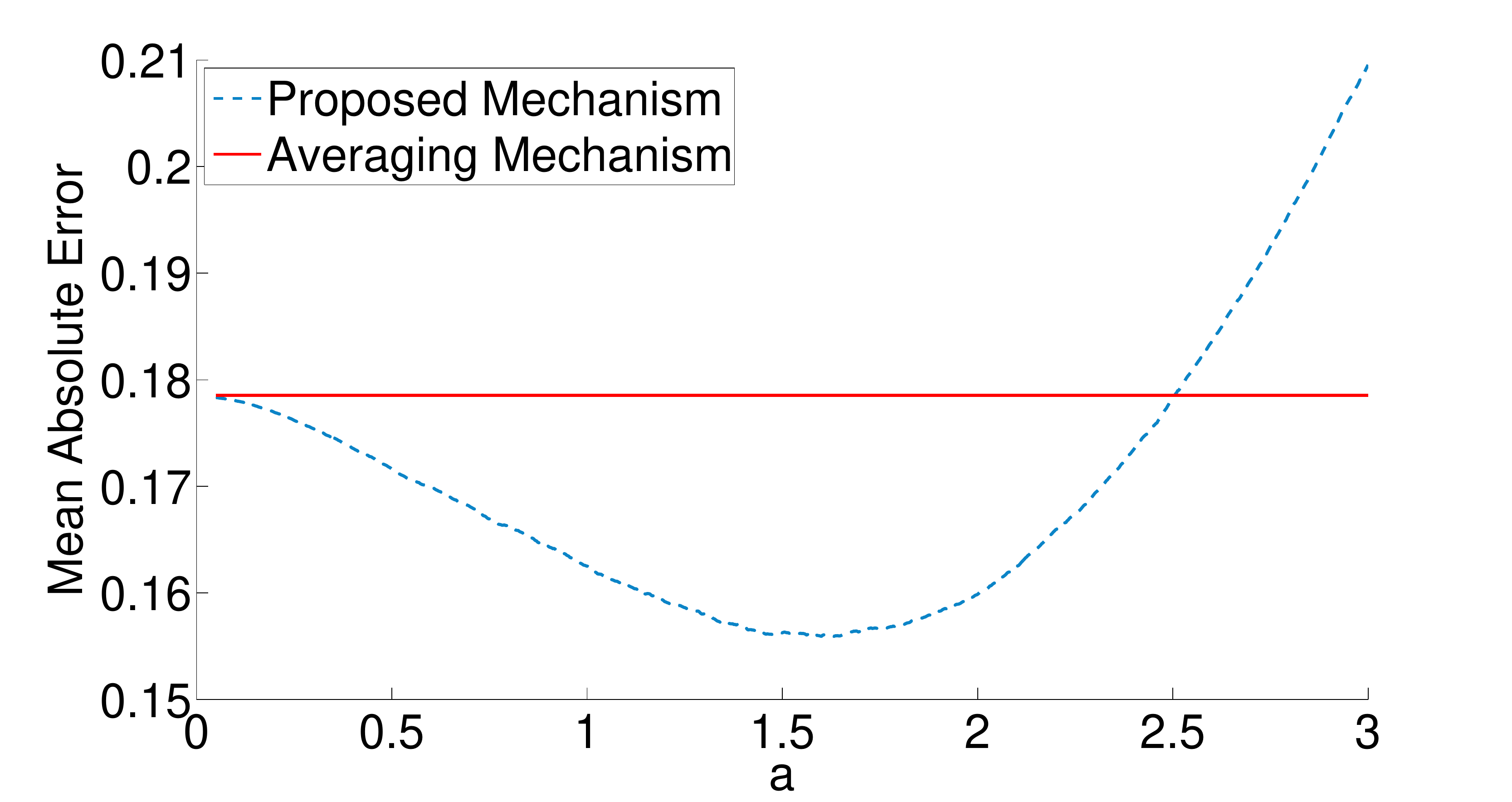}%
\caption{Errors vs. $a$}
\label{MAE_vs_Simple}%
\end{minipage}
\begin{minipage}[b]{0.33\linewidth}
\includegraphics[width=0.8\columnwidth]{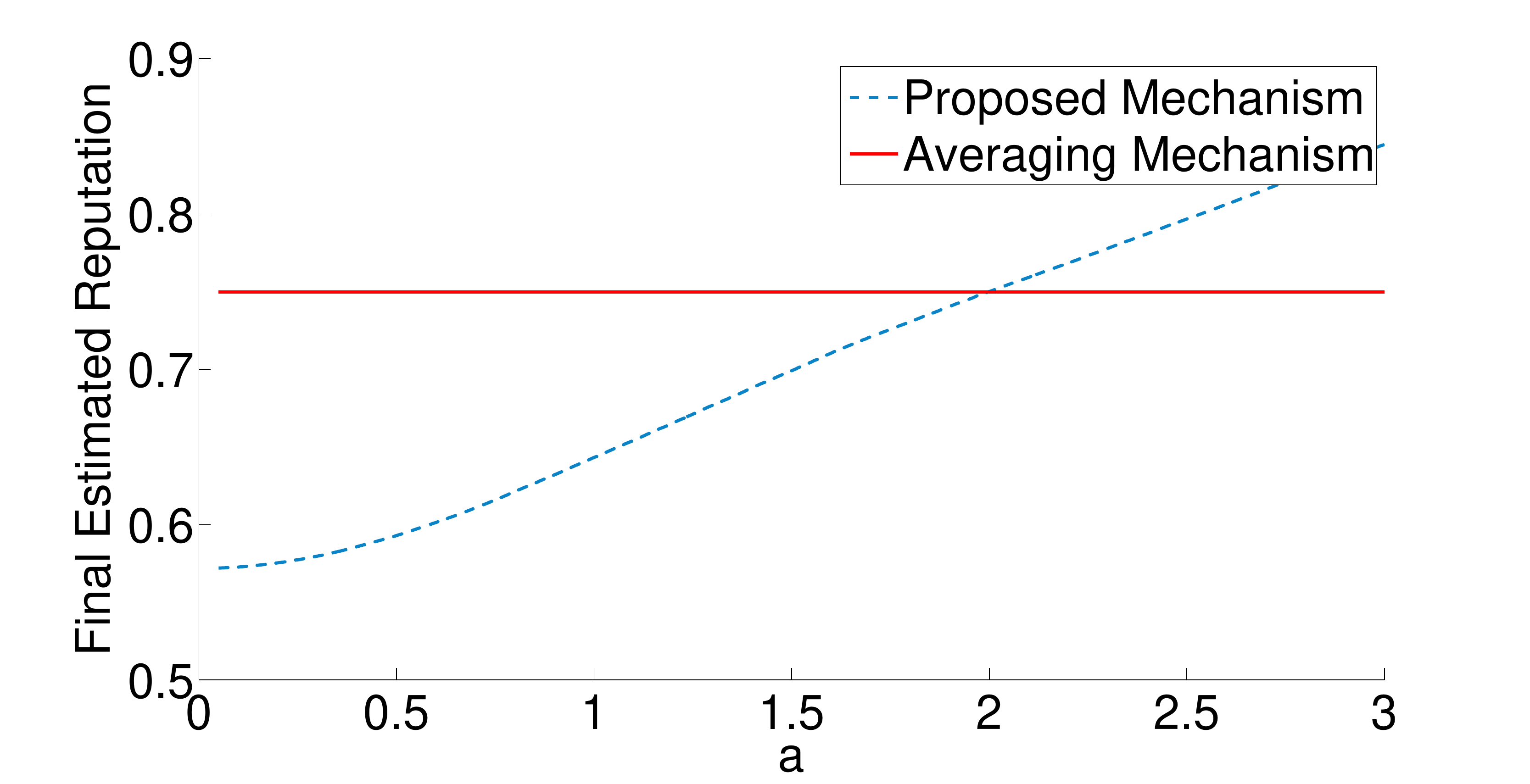}%
\caption{Est. reputation vs. $a$}
\label{Est_vs_a}%
\end{minipage}
\end{figure*}

As seen in \eqref{eq:ErrorDer}, the optimal choice of $a$ does not depend on the specific values of $\mu$ and $\sigma'$. Therefore, the same mechanism can be used for any set of users. Equation \eqref{eq:ErrorDer} can be solved numerically, reflecting that the minimum error is achieved at $a\approx 1.7$.  This can be seen from Fig. \ref{MAE_vs_Simple}, which shows the MAE of the PR mechanism compared to that of the averaging mechanism. Under the simple averaging mechanism the MAE is $E[|\bar{x}_{0i}-r_{ii}|] = \sqrt{\frac{2}{\pi}}\sigma'$.
We see that for a large range of $a$ values the PR mechanism given in \eqref{eq:GameForm} results in smaller estimation error.  This suggests that $N_i$'s self-report can significantly benefit the system as well as all users other than $N_i$. 

We have now verified the PR mechanism as a suboptimal solution to the centralized problem \eqref{eq:sys_goal} under Model III.  It is clearly budget balanced as no taxation is invoked. 
%
%
We next check whether there is incentive for $N_i$ to provide its self-report, i.e., does this benefit $N_i$ itself?  Fig. \ref{Est_vs_a} compares $N_i$'s estimated reputation $\hat{r}^A_i$ under the proposed mechanism to that under the averaging mechanism\footnote{In calculating the reserved utility, we have assumed that in the event $N_i$ chooses to stay out of the game, the reputation system will use simple averaging on the gathered cross-observations to estimate $N_i$'s reputation}, 
in which case it is simply the average of all observations on $N_i$, and $E[\bar{x}_{0i}]=r_{ii}$ when unbiased. 

Taking Figs. \ref{MAE_vs_Simple} and \ref{Est_vs_a} together, we see that there is a region, $a\in [2, 2.5]$ in which the presence of the self-report helps $N_i$ obtain a higher reputation index, while helping the system reduce its estimation error on $N_i$.  This is a region that is mutually beneficial to both $N_i$ and the system, and $N_i$ clearly has an incentive to participate and provide its self-report.

It remains an interesting problem and a challenge to find the mechanism 
that results in the \emph{smallest} performance gap, if it exists, compared to the solution to the centralized problem \eqref{eq:sys_goal}. 

\subsection{An extension to the PR mechanism} 
%

A variation on the preceding PR mechanism would be to use the weighted mean of the cross-reports instead of a simple average:
\begin{eqnarray}
\bar{x}_{0i}:= \frac{\sum_{j\in K\backslash i} {w_j}x_{ji}}{\sum_{j\in K\backslash i} {w_j}}
\label{eq:weighted mean}
\end{eqnarray}
where $\underline{w}:=(w_{j})_{j\in K\backslash i}$ is a vector of weights, also specified by the reputation system. One reasonable choice for $\underline{w}$ could be a vector of previously computed reputations $\hat{r}^A_j$, with the intention of allowing the more reputable users to have a higher influence on the estimates.  Similar ideas are commonly used in rating/ranking systems. 
We proceed by analyzing the performance of this alternative mechanism. 

Assume $x_{ji}\sim N(r_{ii}, \sigma_{ji}^2)$, 
i.e., all users have an unbiased view of $N_i$, but with potentially different accuracy as reflected by different values of $\sigma_{ji}$, with smaller variances corresponding to more precise estimates. 
%
In the special case $\sigma_{ji}=\sigma$, $\forall j$, it can be shown that the weighted average will (regardless of the choice of $\underline{w}$) {\em increase} the variance of the aggregated cross-report, and thus the estimation error.  What this implies is that users with equally accurate views should be given the same power to affect the outcome. 

On the other hand, if $\sigma_{ji}$'s are different for different users, then choosing $\underline{w}$ such that $\sum_{j\in K\backslash i} {w_j}^2\sigma^2_{ji} \leq \sum_{j\in K\backslash i} \frac{1}{(K-1)^2}\sigma^2_{ji}$ results in a lower variance, and thus a lower estimation error. This rearrangement shows clearly that for the inequality to hold, it suffices to put more weight on the smaller $\sigma_{ji}$'s, i.e., more weight on those with more accurate observations.  
%
%
%
Technically this result is to be expected.  However, in our context it points to the following interesting interpretation: more reputable users (higher $\hat{r}^A_{j}$) should only be given higher weights if they also have more accurate observations (smaller $\sigma_{ji}$), which may or may not be the case.  This is a scenario where reputation itself should not carry more voting power.  Otherwise the system is better off assigning equal weights to all. 


%% file: Model_V.tex
\section{A Heterogeneous Scenario} \label{ModelV}
%
%
So far we have only considered homogeneous sets of users.  
In this section, we consider a simple heterogeneous setting: of the $K$ users, $T$ of them are of the truth type, with utility functions given by \eqref{eq:ind_goal_Ia}, while $I=K-T$ are of the image type, 
 with utility functions given by:  
\begin{eqnarray}
\mbox{(Model IV)} & u_i =& g_i(\hat{r}^A_i)~. 
\label{eq:ind_goal_V}
\end{eqnarray} 
Specifically, we study the inefficiency resulting from naively adopting the Absolute Scoring mechanism. 

\subsection{Image type users} \label{imageV}
 
A user $N_i$ in $I$ will choose its self-report so as to achieve $\max_{x_{ii}} E[v_i]$. Solving this optimization problem assuming $R_{0i} \sim {\cal N}(r_{ii}, \sigma^2)$ results in:
\[x^*_{ii} = \{x: g'_i(x) = 2(x - r_{ii}) \}~.\]
As expected, $N_i$'s strategy depends on its valuation of an inflated reputation index, i.e., it depends on the functional form of $g_i(\cdot)$, with the interpretation that $N_i$ will inflate its report as long as the marginal increase in tax payment is no more than the marginal gain from an inflated report. In the special case of $g_i(x)=x$, the optimal self-report  is given by $x_{ii}^*=\min \{r_{ii}+ \frac{1}{2} ,1\}$. 

We next verify the individual rationality condition for these participants. The biased self-report resulting from the optimization problem is given by $x_{ii}^* = r_{ii}+ \frac{g_i'(x^*_{ii})}{2}$. Thus the utility in staying out or participating is, respectively: 
\[U_{i}^{\{Out\}} = E(g_i(R_{0i}))\leq g_i(r_{ii})\quad \text{(by concavity of $g_i(\cdot)$)} \] 
\[U_{i}^{\{In\}} = g_i(x_{ii}^*) - \frac{(g_i'(x_{ii}^*))^2}{4} + \frac{1}{K-1}  \sum_{j\in I, j\neq i}\frac{E[(g_j'(x_{jj}^*))^2]}{4}~.\]  
$N_i$ has incentive to participate if $U_i^{\{in\}}\geq U_i^{\{Out\}}$, a condition highly dependent on the specifics of the system. We will look at one example in detail in \ref{exampleV}.  
%
\subsection{Truth type users} \label{truthV}

It is obvious that a user $N_i$ in $T$ will choose its self-report truthfully, i.e., $x_{ii}^*=r_{ii}$. 
The complication, however, is in ensuring that this user has an incentive to participate. Intuitively, the problem arises from the presence of image type users who introduce inaccuracy in the reputation system, making $N_i$ less interested in (trusting of) the outcome, and consequently less likely to participate. We formalize this intuition as follows.  

First suppose user $N_i$ decides to stay out. If $N_i$ has its own unbiased cross-observations (assuming at no additional cost), the expected utility of $N_i$ from staying out is given by:
\[U_{i}^{\{Out\}} = -\sum_{j\neq i} E[f_i(| R_{ij} - r_{jj} |)] ~.\] 

Next, consider the expected payoff from participation. The simplified expression is given by:\footnote{Assuming all cross-observations have the same variance $\sigma^2$.}  
\begin{eqnarray*}
U_{i}^{\{In\}} &=& -\sum_{j\in I} \Big(E[f_i(| x_{jj}^* - r_{jj} |)] \\
&-& \frac{1}{K-1} E[| x_{jj}^*-R_{0j}|^2] \Big) - \frac{I}{K-1} \sigma^2 ~.
\end{eqnarray*}

A user $N_i$ has an incentive to participate if $U_i^{\{in\}}\geq U_i^{\{Out\}}$, a condition  dependent on the specific $R$, functional forms of $f_i(\cdot)$ and $g_i(\cdot)$, and $\sigma^2$ of the  system.  A concrete example is given in \ref{exampleV}.

\subsection{Effects on the system performance} 
We now consider whether the implementation of {the AS} mechanism in such a heterogeneous environment will improve upon the direct observations of the reputation system.  For comparison, consider a reputation system that simply assigns the reputations $\hat{r}^A_i=R_{0i}$, $\forall i\in K$.} The performance of this mechanism according to \eqref{eq:sys_goal} will be $\sqrt{\tfrac{2}{\pi}}K\sigma$. On the other hand, the performance of the AS mechanism is given by: 
\[\sum_{i\in K} |\hat{r}^A_i - r_{ii}| = \sum_{i\in I} \frac{g_i'(r_{ii})}{2}~.\]
Thus such implementation is profitable if:
\[\sum_{i\in I} g_i'(r_{ii})<2\sqrt{\frac{2}{\pi}}K\sigma~.\]

\subsection{Example: $f_i(x)=x^2$ and $g_i(x)=x$} \label{exampleV}

Recall that the optimal self-report for a user $N_j$ in $I$ is (at most) $x_{jj}^*= r_{jj} + \frac{1}{2}$. 
We can thus simplify the expected utility expressions in \ref{truthV} to get: 
\[U_{i}^{\{Out\}} = -(K-1)\sigma^2 ~.\] 
\[U_{i}^{\{In\}} =  -\frac{I}{4}(1-\frac{1}{K-1})~.\] 

Therefore, $N_i$ has an incentive to participate if: 
\[\frac{I}{K-1} \leq 4\frac{K-1}{K-2}\sigma^2\] 

Define $\rho:=\frac{I}{K-1}$ as the (estimated) fraction of the image type users. 
The following condition is thus sufficient to guarantee voluntary participation by a truth type $N_i$:
\[{\rho\leq 4\sigma^2} ~. \]

This result coincides with our initial intuition: the higher the percentage of image type users (larger $\rho$), the less likely is a truth type user to participate. Also, given a high accuracy in a truth type's own cross-observations (smaller $\sigma^2$), this individual is less interested in participating in a crowd-sourcing mechanism.  

We next check whether the image type users have an incentive to participate. 
Using the expressions obtained in \ref{imageV}, we see that an image type $N_i$ has an incentive to participate if:
\[\max\{r_{ii}+\frac{1}{2}, 1\} - r_{ii} \geq \frac{1}{4}\frac{T}{K-1} ~.\]

It is easy to see that the image type users with $r_{ii}\leq 0.5$ always have the incentive to participate in the mechanism. On the other hand, the participation of higher quality image type users, with $r_{ii}>0.5$, is harder to guarantee. Define $\gamma:=\frac{T}{K-1}$ as the fraction of the truth type users. Such high quality $N_i$ will choose to participate in the mechanism if: 
 \[{\gamma\leq 4(1-r_{ii})} ~.\] 
The intuition behind this result is the following: due to the bias introduced by $N_i$, the expected tax payment of this user is positive, unless there are many other image type participants (small $\gamma$), such that the reallocation of their paid taxes will offset this payment.  The lower the reputation of $N_i$ (smaller $r_{ii}$), the more it hopes to (or the more its potential to) gain by inflating its report  through the proposed mechanism, and therefore it has more incentive for participation.  

Finally, we verify whether implementing the proposed mechanism is reasonable from the viewpoint of the system. The performance of the mechanism under the current specifications is given by:
\[\sum_{i\in K} |\hat{r}^A_i - r_{ii}| = \sum_{i\in I} |\hat{r}^A_i - r_{ii}| < \frac{1}{2}I ~.\]
{Therefore, the following condition is sufficient for the reputation system to gain:}
\[\rho<2\sqrt{\frac{2}{\pi}}\sigma~.\]  
Intuitively, the benefit of the proposed mechanism is decreasing in the accuracy of the estimations (higher $\sigma$), and decreasing in the fraction of image type users. 


%% file: Discussion.tex
\section{Discussion} \label{sec:disc}

\subsection{Other possible environments} \label{otherenv}

We omitted the analysis of a few other possible environments, including homogeneous environments of image type users, and heterogeneous environments of image type and mixed type users. When absolute reputation indices are used, our analysis of the PR mechanism can be easily extended to include such scenarios. More specifically, the impossibility result of Section \ref{ModelIV} continues to hold {in the above two cases}, and a similar PR mechanism may be used to obtain sub-optimal system performance. The challenge in dealing with the presence of image type users lies in the fact that as the fraction of mixed type users (if present) decreases, the available valid cross-observations also decrease. 
Even though as noted in Section \ref{ModelIV} the PR mechanism can operate using only the system's observations, the decreased accuracy of the aggregate cross-report degrades the system performance. 



In general the image element of a utility function introduces additional complexity to the problem even when all users are of the same, mixed type.  
To further illustrate, consider the following homogeneous environment with the mixed-type utility function of proportional indices: 
\begin{eqnarray}
u_i = - \sum_{j\neq i} f_i(| \hat{r}^R_j - \frac{r_{jj}}{\sum_{k} r_{kk}}|) + g_i(\hat{r}^R_i)~.
\label{eq:ind_goal_IIb}
\end{eqnarray}
 This model bears similarity to existing resource allocation problems, see e.g. \cite{auction ,leverage}, with one fundamental difference that has to do with the relationship between the users' utility and the system or global objective. 
{In the existing literature, the global objective of the centralized allocation problem (or the social choice rule) is often taken to be the sum of individual utilities. The desired outcome is then induced by aligning individual users' objectives with the social choice rule using taxation.} All allocation mechanisms that are based on the Vickrey-Clarke-Groves (VCG) \cite{micro, game} mechanism are examples of this approach, see e.g., \cite{leverage}. 
For the utility function in \eqref{eq:ind_goal_IIb}, 
this approach is not applicable due to the presence of the extra term $g_i(\hat{r}^R_i)$. 

More precisely, consider a direct mechanism where all users $N_j$ disclose $r_{jj}$ as part of their message space, but $N_i$ has unilaterally deviated to reporting $x_{ii}\neq r_{ii}$.
Let's see how $N_i$'s utility changes when deviating:
\begin{eqnarray}
&& v_i(x_{ii},\{r_{kk}\}_{k\neq i}) - v_i(\{r_{kk}\}_{k=1}^K)=\notag\\
&& ~~ - \left(\sum_{j\neq i} f_i(|\tfrac{r_{jj}(x_{ii}-r_{ii})}{(x_{ii}+\sum_{k\neq i} r_{kk})(\sum_k r_{kk})}|) - \sum_{j\neq i} f_i(0)) \right) \notag\\
&& ~~ + \left( g_i(\tfrac{x_{ii}}{x_{ii}+\sum_{k\neq i} r_{kk}}) - g_i(\tfrac{r_{ii}}{\sum_{k} r_{kk}}) \right)\notag\\
&& ~~ - \left(t(x_{ii})-t({r_{ii}})\right) ~.
\label{eq:profit}
\end{eqnarray}  
In \eqref{eq:profit}, the first term is always negative since $f_i(0)\geq 0$ and increasing; it represents the loss incurred by the inaccuracy that a false report introduces to the system. The second term is positive for $x_{ii}>r_{ii}$ and negative for $x_{ii}<r_{ii}$. {This term captures the profit from an inflated report $x_{ii}$ and thus an inflated $\hat{r}^R_i$.}   The last term, which can be used if required, is the difference between the tax paid by $N_i$ in the two cases. 

Depending on the functions $f_i(\cdot)$ and $g_i(\cdot)$, two scenarios are possible {regardless of the form of the taxation}. 

\emph{Case I}:	These functions are such that the benefit from increased $\hat{r}^R_{i}$ is not worth the loss in the system accuracy. In this case the users will not deviate from truth-telling and therefore the centralized solution is implementable. For example, it can be shown that the centralized solution is implementable in the special case of $f_i(|x|)=|x|$ and $g_i(x)=x$, using the Absolute Scoring mechanism but with a proportional allocation rule. 

 \emph{Case II}: There is a net benefit resulting from the first two terms in \eqref{eq:profit}.  In this case the tax terms can be chosen so as to ensure truth-telling, and the individual rationality and budget balance constraints have to be carefully addressed which can be done for specific choices of $f_i(\cdot)$ and $g_i(\cdot)$.  

\subsection{Taxation: interpretation \& implementation}\label{sec:tax}

The AS mechanism introduced in Section \ref{ModelI} relies on taxation to induce truth revelation.  
It finds natural interpretation and plausible implementation in both our application contexts.  In the case of online shopping, the shopping site runs the reputation system and can implement the taxes as a form of user (seller and buyer)-specific transaction/service surcharge or credit.  Again as the mechanism is budget balanced, the site retains no money from this surcharge; it is merely a means to induce desired behavior by reallocating the money among users. 
In the case of network reputation, taxation may be implemented in a number of ways.  One way is that 
 networks might charge each other a premium (or give each other credit) for access to content depending on the taxation amount. 
Another method by which taxation might be implemented is in the negotiation of peering policies.  

\subsection{Different types of leverage and the effect of externalities} \label{leverage}


As mentioned in Section \ref{sec:MD}, mechanism design typically relies on some form of leverage to induce desired behavior, taxation being a very common one.  Below we identify a few other factors {\em inherent} in the model that serve as leverage our mechanisms take advantage of.  The first concerns the difference between the absolute and proportional indices.  With absolute indices, unilateral deviation to inflate one's reputation does not result in loss of accuracy in estimates of other users, while with proportional indices the increased index comes at the cost of accuracy to the system or the user itself. As a result proportional reputation carries leverage for the system when combined with the truth element; this is seen in the case of all truth type users (Section \ref{ModelII}, where under the FR mechanism taxation is not needed due to sufficient leverage introduced by proportional indexing compared to absolute indexing), and in the case of all mixed type users (Section \ref{otherenv}, where implementation of the centralized solution is feasible using taxation compared to the impossibility result for absolute indexing). 

{ To further illustrate the difference between absolute and proportional indices, we point out that the problems studied herein, for both the absolute and proportional reputation settings, bear resemblance to \emph{public good} problems, in that the vector of reputations affects the utility of all users simultaneously. It is known that uncoordinated markets result in inefficiency in the provision of a public good \cite{micro}. Consequently, any proposed mechanism has to ``internalize the externalities'' in order to provide the optimal level of public good, here the reputation indices. To this end, the AS mechanism requires users to pay according to the negative externality they impose on others due to their inaccurate self-reports. On the other hand, the use of proportional reputations together with the truth element in users' utilities automatically internalizes the externalities, removing the need for additional taxation (the FR mechanism). 
} 
 
Any independent observation the reputation system possesses, and the public knowledge of its possession of such information, can also serve as a powerful threat to the users to not deviate, leading to simpler mechanisms and better system performance.  It is in particular indispensable in dealing with image type users, as pointed out in \ref{otherenv}.  The presence of the truth element in users' utilities may simplify a mechanism by removing the need for independent observations. 

\subsection{Malicious users} 

Throughout Sections \ref{ModelI}-\ref{ModelV}, in both homogeneous and heterogeneous scenarios, we focused on users whose utility function can be described by the truth and/or image elements introduced in Section \ref{sec:model}. These elements do not necessarily capture the strategic behavior of \emph{malicious} users, whose utility functions, and thus their strategic reports, may or may not be known to the reputation system.  Without introducing a concrete rational model for a malicious user, here we briefly discuss the effect of {\em random} reporting as a form of malicious behavior (mischievous is perhaps a better word). 

\emph{Random self-reports:} We consider a heterogeneous scenario consisting of malicious users and truth type users, where the Absolute Scoring mechanism is used. This analysis will in turn allow for a comparison between the effects of random participation and the targeted strategic behavior of the image type (Section \ref{ModelV}). We can show that in many scenarios, including the example in \ref{exampleV}, image type users are more detrimental to the system's accuracy as compared to malicious users with random reports, since they are specifically choosing self-reports that cause inaccuracy in the system. 

\emph{Random cross-reports:} In reputation systems that rely on cross-observations (e.g. the PR mechanism), random cross-reports pose a different dilemma for the reputation system. While the increased number of cross-reports increases the accuracy of the reputation mechanism, collecting random reports in this process will inevitably hurt the performance by degrading the quality of the aggregated cross-observation. 

\subsection{Dealing with Collusion} \label{collusion}

We now briefly discuss the handling of collusion/cliques.  Without establishing a formal way of analyzing collusion in this context, we note that our mechanisms possess features that can naturally prevent or reduce the effect of collusion.  

	The Absolute Scoring and Fair Ranking mechanisms (Sections \ref{ModelI} and \ref{ModelII}) are naturally collusion-proof: they either do not require cross-reports, or use  the system's independent observations to incentivize truthful input. Consequently, collusion cannot be used to inflate one's own reputations or degrading others'. 
			
	Consider now the Punish Reward (PR) mechanism (Section \ref{ModelIV}). 
	Participation of clique members who provide unfair high/low cross-reports may disrupt the performance of this mechanism (which in the absence of cliques outperforms the simple averaging mechanism, the mechanism currently used at eBay). However, the PR mechanism can remain functional using only the cross-observations from a subset of trusted entities, or even with a single observation by the reputation system. The PR mechanism with one or limited truthful input can prevent both slandering and promoting attacks. Also, given enough truthful input, PR results in a more accurate outcome than simply averaging the cross-observations from all users, some of which may be part of a clique. 

	{ Finally, if the system lacks independent observations, introducing randomness in the mechanisms can reduce the impact of cliques. To illustrate, consider the extended-AS mechanism (Section \ref{ModelI}), where each user is charged according to the inaccuracy of its self-report when compared to a randomly selected peer. The extended-AS in its current form is vulnerable to collusion, since members of a clique can attack the mechanism by agreeing on inflating each other's reputations, or by bashing other users to extract revenue from the redistributed taxes.  

To reduce the profitability of forming cliques, we can impose additional layers of taxation (not necessarily more taxes) to the extended-AS mechanism. We illustrate this idea using two layers of taxation. Assume a user $N_i$ is further charged based on the discrepancy between its cross-report $x_{i(i+1)}$ and its predecessor's cross-report $x_{(i-1)(i+1)}$: 
\begin{eqnarray*}
t_i = | x_{ii}-x_{(i-1)i}| - \tfrac{1}{K-2} \sum_{j \neq i, i+1} | x_{jj}-x_{(j-1)j}|~~~~~~~~~~~~~~~\notag\\
		+ | x_{i(i+1)}-x_{(i-1)(i+1)}| - \tfrac{1}{K-2} \sum_{j \neq i, i+1} |x_{j(j+1)}-x_{(j-1)(j+1)}|.
\end{eqnarray*} 
The following proposition verifies that truthful cross-reports are best-responses to unbiased cross-reports. 

\begin{proposition}
When $x_{(i-1)(i+1)}$ is truthful, $| x_{i(i+1)}-x_{(i-1)(i+1)}|$ is minimized with a truthful cross-report $x_{i(i+1)}=R_{i(i+1)}$. 
\end{proposition}
\begin{IEEEproof}
Define $Z:=| x_{i(i+1)}-x_{(i-1)(i+1)}|$. We use known results on folded Normal distributions. If $X\sim {\cal N}(\mu, \sigma^2)$, the random variable $Y = |X|$ has a folded Normal distribution, the expected value of which is given by: 
\[E[Y]=\sigma\sqrt{2/\pi}\exp{(-\mu^2/2\sigma^2)} + \mu(1-2\Phi(-\mu/\sigma))~, \]
where $\Phi(\cdot)$ denotes the CDF of the standard Normal distribution. 

Assume $x_{(i-1)(i+1)}\sim {\cal N}(r_{i+1}, \sigma^2)$ is truthful, and that $N_i$ manipulates its cross-report such that $x_{i(i+1)}=aR_{i(i+1)}+b \sim {\cal N}(ar_{i+1}+b, a^2\sigma^2)$. Then: 
\[E[Z]=\hat{\sigma}\sqrt{2/\pi}\exp{(-\hat{\mu}^2/2\hat{\sigma}^2)} + \hat{\mu}(1-2\Phi(-\hat{\mu}/\hat{\sigma}))~, \]
where $\hat{\mu}:=(a-1)r_{i+1}+b$ and $\hat{\sigma}^2=(1+a^2)\sigma^2$. This expression can be further simplified to: 
\[E[Z]=\hat{\sigma}\sqrt{2/\pi}\exp{(-\hat{\mu}^2/2\hat{\sigma}^2)} + \hat{\mu}\text{ erf}(\hat{\mu}/(\sqrt{2}\hat{\sigma}))~, \] 
$E[Z]$ will be minimized with $a=1$ and $b=r_{i+1}$. Therefore, $N_i$'s best-response is $x_{i(i+1)}=R_{i(i+1)}$. 
\end{IEEEproof}

We note that to increase the randomness in the mechanisms, and consequently decrease the predictability of being matched with clique users, each layer of taxation can be assessed according to a different, undisclosed random ring. This will help further strengthen the extended-AS mechanism against collusions. An increased likelihood of being matched with honest/non-clique users will reduce the benefit of forming cliques. Thus cliques will have limited benefit unless their size is comparable to the user population.

\subsection{Consistency with Existing Impossibility Results} 
Impossibility results in mechanism design specify combinations of properties that no mechanism can satisfy simultaneously. The basic approach for establishing such results is through the revelation principle \cite{parkes01b}, which states that if a (possibly indirect) mechanism implements a social choice function in dominant strategies, then there exists a direct incentive compatible mechanism that implements the same social choice function in dominant strategies (similarly for BNE). Inconsistencies can then be found by mathematically checking incentive compatibility, individual rationality, and other desired properties, in a direct mechanism. 
%
Each impossibility result is derived for a certain solution concept, preference type, and environment. Impossibility results for weaker solution concepts, restricted utility types, and restricted environments are thus stronger, since they include more general settings as special cases \cite{parkes01b}.

One of the very first impossibility results is that of Gibbard \cite{gibbard73}, which states that when the space of players' utilities is sufficiently rich, a social choice function is implementable in dominant strategies if and only if it is the trivial dictatorial rule. This is a rather weak impossibility result, as it encompasses general user preferences and general environments; it is applicable for instance to voting mechanisms where users are allowed to have general preferences. \cite{gibbard73} established the conjecture that in such general settings, no voting system is immune to strategic voting. However, in most engineered systems the preference environments are more structured, such that implementation of non-dictatorial social choice functions is often possible.  
	
	A stronger version of Gibbard's impossibility result is that of Green and Laffont \cite{green77}. The authors show that in environments where the decisions are on the level of a public good and the amount of individual transfers, and the agents' utilities are separable in income (i.e. quasi-linear) with general valuation functions, the only direct mechanisms that implement an allocatively efficient outcome in dominant strategies are Groves (VCG) mechanisms. Consequently, since VCG mechanisms are only \emph{weakly} budget balanced, it is impossible to achieve efficiency and (strong) budget balance even in quasi-linear environments. 
	
	Green and Laffont's result is further strengthened by relaxing the solution concept to Bayesian Nash equilibrium. Myerson \cite{myerson83} shows that in simple-exchange environments \cite{parkes01b} with quasi-linear preferences, it is impossible to achieve efficiency, budget balance, and individual rationality in a Bayesian Nash incentive compatible mechanism. 
	
	A more recent result by Jackson \cite{jackson91} closes the gap between necessary and sufficient conditions for Bayesian implementation in economic environments (including environments in which there are public goods and/or externalities). In such environments with 3 or more players, a social choice function is implementable in a Bayesian Nash equilibrium if and only if incentive compatibility, closure, and Bayesian monotonicity conditions are satisfied. 
	

	At first sight, these impossibility results (especially \cite{green77, myerson83}) may seem contradictory to our efficient, individually rational, budget balanced, direct AS and FR  mechanisms. It should be noted however that we are restricting attention to a specific class of valuation functions (rather than quasi-linear utilities with general valuation functions). Therefore we are able to find non-VCG mechanisms achieving the aforementioned properties. On the other hand, the impossibility result of \cite{jackson91} is indeed applicable in Section \ref{ModelIV}, which justifies the search for suboptimal mechanism, such as the PR mechanism. 
 

%% file: Related.tex
\section{Related Work} \label{sec:related} 

The theory of mechanism design, originally proposed for problems in the economic literature, has been increasingly used to address problems of resource allocation in informationally decentralized systems with strategic users. Pricing schemes, e.g. \cite{pricing}, and auctions, e.g. \cite{auction}, are two popular approaches in the design of allocation schemes in communication systems.  The use of pricing allows the system to align individual users' objectives with global performance goals to implement socially optimal outcomes.  Taxation may be viewed as a form of pricing, which we have used in Section \ref{ModelI}. 

Despite the feasibility of using monetary taxation in our setting (see Section \ref{sec:tax}), alternative forms of leverage are usually preferred in incentivizing user cooperation though they are relatively hard to identify; two notable exceptions are \cite{leverage, intervention}. 
In \cite{leverage}, the authors study the problem of using the downlink rate allocated to a user as an alternative commodity to induce socially optimal uplink rate allocation in a multi-access broadcast channel with selfish users,  
%
%
while \cite{intervention} proposes an \emph{intervention} mechanism  
that uses the \emph{commodity of interest} as the means for preventing users' deviation from their designated strategies. Specifically, a monitoring device is used to estimate the transmit power profile of selfish users in a wireless network; it then chooses to transmit at a positive power level if users deviate to higher transmission powers, thus negatively affecting users' utilities.  
The PR mechanism in Section \ref{ModelIV} also relies on a credible threat of punishment to deter non-cooperative users from deviation. 
However, the above intervention mechanism can only exercise punishment while our PR mechanism can also reward users' cooperative actions using the commodity of interest.  

The work presented in this paper is also closely related to elicitation and prediction mechanisms used for aggregating the predictions of agents about an event, see e.g. \cite{scoring, serum, col_rev}. Scoring rules \cite{scoring} incentivize an agent to truthfully reveal its prediction by offering rewards based on the accuracy of the agent's estimation as compared to the actual realization of the event. Although these rules can be used to quantify the performance of forecasters, they rely on the observation of an objective ground truth. A class of \emph{peer prediction} methods can be used to eliminate the need for such verification by requesting an agent's own assessment, as well as its prediction of other agent's assessment. For example, the elicitation methods in \cite{serum} and \cite {col_rev} result in truthful revelation even for subjective assessments. 
However, in all aforementioned works, the users are essentially rewarded in accordance with their participation, but do not attach any value to the realization of the event, or the outcome that the elicitor may be building using the aggregated data. Among our proposed models, Model I resembles this line of work due to the absence of an image element in users' utilities. Consequently, users do not attach value to the outcome that is built using their inputs, i.e., the reputation index derived from their self-report. The AS mechanism is nevertheless different, in that users receive non-monetary rewards (a vector of accurate reputation indices) by participation, whereas in elicitation methods monetary rewards are used to incetivize cooperation. Furthermore, although we are studying a problem of elicitation about an objective ground truth, this event is not observable by the elicitor.

There has also been a large volume of literature on the use of reputation in peer-to-peer (P2P) systems and other related social network settings, including but not limited to, blogs, forums \cite{josurvey}, and corporate wikis \cite{dencheva11} that depend heavily on user contribution, opportunistic forwarding networks \cite{bigwood11}, trust management in ad-hoc networks \cite{buchegger02}, and the like. 
Specifically, a large population, the anonymity of individuals in such social settings, and the lack of proper rewards make it difficult to sustain cooperative behavior among self-interested individuals \cite{P2P_1,josurvey}.    
Reputation has thus been used in such systems as an incentive mechanism for individuals to {\em cooperate} and behave according to a certain {\em social norm} in general \cite{social_norm,josurvey}, and to reciprocate in P2P systems in particular. 
While the focus of social network studies is on the effect of changing reputation on individuals, the focus of our study is on how to make reputations an accurate representation of a user's quality. 
Accordingly, our emphasis is on how to incentivize participation from users, and whether the system could obtain the true quality of the users. 
Furthermore, observations in our system are noisy and incomplete, while they are typically assumed to be perfect in P2P systems as they are based on direct or indirect personal/social reciprocation \cite{reputation-P2P_1,reputation-P2P_2}. Finally, we have aimed to \emph{incentivize} truthful cross-observations, while the schemes in \cite{bigwood11, dencheva11, buchegger02} either rely on external verification mechanisms to \emph{enforce} truthful reports, or do not explicitly address the truthfulness of such reports.  

In addition to incentivizing participation from \emph{all} users involved in an interaction, another main feature of our proposed mechanisms is the balance between transparency and robustness: a simple reputation system provides clarity for users to decide about participation and allows easy interpretation of results, while sufficient built-in robustness prevents the system from being manipulated by strategic users. This approach is different from the commonly used idea of ``security by obscurity'', i.e., keeping (some parameters of) a sophisticated reputation system confidential so as to hinder manipulation and maintain robustness, which has been advocated by J{\o}sang et al. in \cite{josurvey}, and is currently being used in various reputation systems, including Amazon's reviewers' rankings, Google's PageRank, and the product review site Epinions \cite{josurvey}. 


%% file: main_reputation.bbl
\begin{thebibliography}{10}

\bibitem{dnssensor}
M.~Antonakakis, R.~Perdisci, D.~Dagon, W.~Lee, and N.~Feamster.
\newblock {Building a Dynamic Reputation System for DNS}.
\newblock In {\em 19th USENIX Security Symposium}, August 2010.

\bibitem{darknet}
M.~Bailey, E.~Cooke, A.~Myrick, and S.Sinha.
\newblock {Practical Darknet Measurement}.
\newblock In {\em 40th Annual Conference on Information Sciences and Systems},
  March 2006.

\bibitem{auction}
J.~H.~R. Berry and M.~Honig.
\newblock Auction-based spectrum sharing.
\newblock In {\em Mobile Networks and Applications 11}, pages 405--418, 2006.

\bibitem{bigwood11}
G.~Bigwood and T.~Henderson.
\newblock Ironman: Using social networks to add incentives and reputation to
  opportunistic networks.
\newblock In {\em IEEE third international conference on Privacy, Security,
  Risk and Trust (passat) and IEEE third international conference on social
  computing (socialcom)}, pages 65--72. IEEE, 2011.

\bibitem{buchegger02}
S.~Buchegger and J.-Y. Le~Boudec.
\newblock Nodes bearing grudges: Towards routing security, fairness, and
  robustness in mobile ad hoc networks.
\newblock In {\em Proceedings of the 10th Euromicro Workshop on Parallel,
  Distributed and Network-based Processing}, pages 403--410. IEEE, 2002.

\bibitem{dencheva11}
S.~Dencheva, C.~R. Prause, and W.~Prinz.
\newblock Dynamic self-moderation in a corporate wiki to improve participation
  and contribution quality.
\newblock In {\em ECSCW 2011: Proceedings of the 12th European Conference on
  Computer Supported Cooperative Work, 24-28 September 2011, Aarhus Denmark},
  pages 1--20. Springer, 2011.

\bibitem{firewalllogs}
DShield.
\newblock {How To Submit Your Firewall Logs To DShield}.
\newblock September 2011.
\newblock \url{http://isc.sans.edu/howto.html}.

\bibitem{reputation-P2P_1}
M.~Feldman, K.~Lai, I.~Stoica, and J.~Chuang.
\newblock Robust incentive techniques for peer-to-peer networks.
\newblock In {\em ACM Conference on Electronic Commerce}, pages 102--111, 2004.

\bibitem{game}
D.~Fudenberg and J.~Tirole.
\newblock {\em Game Theory}.
\newblock The MIT Press, 1991.

\bibitem{gibbard73}
A.~Gibbard.
\newblock Manipulation of voting schemes: a general result.
\newblock {\em Econometrica: Journal of the Econometric Society}, pages
  587--601, 1973.

\bibitem{col_rev}
S.~Goel, D.~Reeves, and D.~Pennock.
\newblock Collective revelation: A mechanism for self-verified, weighted, and
  truthful predictions.
\newblock In {\em Proc. of the 10th ACM Conference on Electronic Commerce},
  2009.

\bibitem{green77}
J.~Green and J.-J. Laffont.
\newblock Characterization of satisfactory mechanisms for the revelation of
  preferences for public goods.
\newblock {\em Econometrica: Journal of the Econometric Society}, pages
  427--438, 1977.

\bibitem{P2P_1}
N.~Hanaki, A.~Peterhansl, P.~Dodds, and D.~Watts.
\newblock Cooperation in evolving social networks.
\newblock {\em Management Science}, 53(7):1036--1050, 2007.

\bibitem{spamcop}
C.~S. Inc.
\newblock {SpamCop Blocking List - SCBL}.
\newblock May 2011.
\newblock \url{http://www.spamcop.net/}.

\bibitem{jackson91}
M.~O. Jackson.
\newblock Bayesian implementation.
\newblock {\em Econometrica: Journal of the Econometric Society}, pages
  461--477, 1991.

\bibitem{josurvey}
A.~J{\o}sang, R.~Ismail, and C.~Boyd.
\newblock A survey of trust and reputation systems for online service
  provision.
\newblock {\em Decision support systems}, 43(2):618--644, 2007.

\bibitem{pricing}
J.~MacKie-Mason and H.~Varian.
\newblock Pricing congestible network resources.
\newblock {\em IEEE Journal On Selected Areas in Communications}, 13(7), Sept.
  1995.

\bibitem{micro}
A.~Mas-Colell, M.~Whinston, and J.~Green.
\newblock {\em Microeconomic Theory}.
\newblock Oxford University Press, 2002.

\bibitem{myerson83}
R.~B. Myerson.
\newblock Mechanism design by an informed principal.
\newblock {\em Econometrica: Journal of the Econometric Society}, pages
  1767--1797, 1983.

\bibitem{parkes01b}
D.~C. Parkes.
\newblock {\em Iterative Combinatorial Auctions: Achieving Economic and
  Computational Efficiency}.
\newblock PhD thesis, Department of Computer and Information Science,
  University of Pennsylvania, May 2001.

\bibitem{serum}
D.~Prelec.
\newblock A bayesian truth serum for subjective data.
\newblock {\em Science}, 306:462--466, 2004.

\bibitem{leverage}
J.~Price and T.~Javidi.
\newblock Leveraging downlink for efficient uplink allocation in a single-hop
  wireless network.
\newblock {\em IEEE Transactions on Information Theory}, 53(11):4330--4339,
  Nov. 2007.

\bibitem{spamhaus}
T.~S. project.
\newblock {SBL, XBL, PBL, ZEN Lists}.
\newblock May 2011.
\newblock \url{http://www.spamhaus.org/}.

\bibitem{reputation-P2P_2}
A.~Ravoaja and E.~Anceaume.
\newblock Storm: A secure overlay for p2p reputation management.
\newblock In {\em International Conference on Self-Adaptive and Self-Organizing
  Systems}, pages 247--256, 2007.

\bibitem{ebay02}
P.~Resnick and R.~Zeckhauser.
\newblock Trust among strangers in internet transactions: Empirical analysis of
  ebay's reputation system.
\newblock {\em Advances in applied microeconomics}, 11:127--157, 2002.

\bibitem{scoring}
R.~Winkler et~al.
\newblock Scoring rules and the evaluation of probabilities.
\newblock {\em TEST/Springer}, 5(1):1--60, 1996.

\bibitem{intervention}
Y.~Xiao, J.~Park, and M.~van~der Schaar.
\newblock Design and analysis of intervention mechanisms in power control
  games.
\newblock In {\em {IEEE} Globecom}, 2011.

\bibitem{social_norm}
Y.~Zhang and M.~van~der Schaar.
\newblock Peer-to-peer multimedia sharing based on social norms.
\newblock {\em Signal Processing: Image Communication}, 27(5):383--400, 2012.

\end{thebibliography}
